\documentclass[aps,twocolumn,groupedaddress,showpacs,notitlepage,nofootinbib]{revtex4-1}
\usepackage{graphicx}
\usepackage{dcolumn}
\usepackage{bm}
\usepackage{amsmath}
\usepackage{amsthm}
\usepackage{multirow}
\usepackage{enumerate}
\usepackage{amsfonts}
\usepackage{ifthen}
\usepackage{psfrag}
\usepackage{slashed}
\usepackage{gensymb}
\usepackage[utf8]{inputenc}
\usepackage{color}
\usepackage{subfigure}
\usepackage{ulem}
\usepackage{amssymb}
\usepackage[colorlinks, citecolor=blue, linkcolor=red, urlcolor=blue]{hyperref}

\newcommand{\be}{\begin{equation}}
\newcommand{\ee}{\end{equation}}
\newcommand{\ber}{\begin{eqnarray}}
\newcommand{\eer}{\end{eqnarray}}

\begin{document} 

\title{The Self-Organized Criticality of Dark Matter in the Early Universe}

\author{Mingjie Jin}
\email{jinmj507@gmail.com}
\author{Ying Li}
\email{liying@ytu.edu.cn}

\affiliation{$Department~of~Physics, Yantai~University, Yantai~264005,~China$ }

\begin{abstract}
We propose a new mechanism for dark matter freeze-out that results in the self-organized criticality in dark matter production, wherein the final relic abundance is independent of initial inputs, in analogy to scale invariance in other realms of non-equilibrium physics. The dynamic of self-organized is triggered through a semi-annihilation process $\chi \phi \to \phi \phi$ in the premise of the instability of dark partner $\phi$, where $\chi$ is the dark matter candidate. The relic abundance can be analytically ascertained when the dark partner is slightly heavier than the dark matter, which permits a substantially heavy dark matter mass without violating unitarity bounds. We demonstrate that this process provides a bridge between the freeze-in and freeze-out mechanisms, and that the intricate dynamics of self-organized criticality can be actualized in a thermal dark sector bath.  
\end{abstract}

\maketitle


{\it Introduction.} Although a plethora of astrophysical and cosmological observations substantiate the existence of dark matter (DM), the intrinsic nature of this substance remains enigmatic. The predominant paradigm assumes that DMs interact with standard model (SM) particles and undergoes a thermal freeze-out from equilibrium when the interaction rate falls below Hubble constant in the early Universe. This has been extensively discussed in the context of weakly interacting massive particles (WIMPs)~\cite{Kolb:1990vq, Jungman:1995df} (see very recent review~\cite{Cirelli:2024ssz}). Additionally, an array of models diverge from the standard paradigm by exploring different thermal freeze-out historical pathways~\cite{ Griest:1990kh, Carlson:1992fn, Finkbeiner:2007kk, Pospelov:2007mp, Feng:2008mu, Hambye:2008bq, Pospelov:2008zw, DEramo:2010keq, Belanger:2012vp, Tulin:2012uq, Hochberg:2014dra, Ko:2014nha, DAgnolo:2015ujb, Kuflik:2015isi, Choi:2016hid, Pappadopulo:2016pkp, Farina:2016llk, Kopp:2016yji, Cai:2016hne, Cline:2017tka, Berlin:2017ife, DAgnolo:2017dbv, Garny:2017rxs, DAgnolo:2018wcn, Kim:2019udq, DAgnolo:2019zkf, Smirnov:2020zwf, Kramer:2020sbb, Bringmann:2020mgx, Erickcek:2020wzd, Heimersheim:2020aoc, DAgnolo:2020mpt, Hryczuk:2021qtz, Fitzpatrick:2020vba, Frumkin:2022ror, Ghosh:2022asg, Bhatia:2023yux}. The current DM relic abundance is linked to DM mass ($m_{\chi}$) as $Y_0 \simeq 0.44\times 10^{-9}\, ({\rm GeV} /m_{\chi})$ at density $\Omega h^2$= 0.12~\cite{Planck:2018vyg}. Clearly, for DM with a typical GeV scale mass, the final relic abundance is significantly lower than initial thermal equilibrium one, which is above $10^{-3}$. The intermediate regime of DM abundance, which lies between these two ends and is produced through a (non-)thermal freeze-out process, has received minimal attention in the literature~\cite{Cheung:2010gj,Du:2021jcj}.

Generally, DM with negligible initial abundance generated through non-thermal processes is described by a freeze-in mechanism, in which DM originates from thermal SM bath yet fails to reach equilibrium due to feeble interaction~\cite{McDonald:2001vt, Hall:2009bx, Chu:2011be, Falkowski:2017uya, Belanger:2020npe, Bernal:2020gzm, March-Russell:2020nun, Bringmann:2021tjr, Boddy:2024vgt, Cervantes:2024ipg}. In fact, non-thermal equilibrium dynamics are a common source of diverse phenomena within physics. Self-organised criticality (SOC) serves as a paradigmatic example for non-equilibrium systems with spatially complex patterns, illustrating a critical state that exhibits scale invariant properties over a wide range of initial conditions or parameters~\cite{PhysRevLett.84.6114, Bertschinger2004RealTimeCA, Kinouchi2006OptimalDR, Berges:2008wm, Schmied:2018mte, PhysRevLett.104.077202, helmrich2020signatures}. The sandpile avalanche is the most well known illustration of the SOC, such as the Bak-Tang-Wiesenfeld (BTW) and Manna model ~\cite{bak1987self,manna1991two}. The SOC is characterized by driven-dissipative system and explained by a mean-field approach~\cite{tang1988mean, Vespignani1996OrderPA, Vespignani1997HowSC, Dickman1997SelforganizedCA, Vespignani1998DRIVINGCA, Hinrichsen2000NonequilibriumCP, Henkel2009NonEquilibriumPT}, i.e., mapping the SOC system onto an absorbing phase transitions. Despite it is not fully understood and remains controversy~\cite{Bonachela2009SelforganizationWC,Watkins2015YearsOS}, the investigations of SOC have been extensive in many aspects, including but not limited to geophysical events like earthquakes~\cite{Sornette1989SelfOrganizedCA}, forest fire~\cite{PhysRevLett.69.1629, Malamud1998ForestFA}, solar flares~\cite{PhysRevLett.96.051102}, complex neuronal activity~\cite{Hesse2014SelforganizedCA}, the Black holes~\cite{Mocanu:2012fd}, the Higgs field~\cite{Eroncel:2018dkg} and cosmic inflation~\cite{Giudice:2021viw}.

In this Letter, we investigate a non-equilibrium dynamics of driven-dissipative ensemble, focusing on weak interaction between DM and unstable dark partner (DP). In this context, we propose a novel DM production mechanism, where a DM particle $\chi$ semi-annihilates with a slightly heavier DP $\phi$, into a pair of $\phi$, followed by $\phi$ {\it irreversible} decaying into both $\chi$ field and an auxiliary $\xi$ field. This process can evolve towards the SOC, facilitated by the slow decay of DP. We are interested in scenarios where the DM exhibits characteristic of the SOC and where the initial abundance of DM is arbitrary, ranging from thermal density to values significantly lower than that of the observed relic abundance. This implies that both DM and DP possess some initial abundances in very early time, prior to the era dominated by SOC. Next, to elucidate the properties of the SOC and their applicability to the evolution of DM, we will begin by discussing some outcomes from the SOC system of Rydberg gas.


\, 

{\it From the Rydberg gas to dark matter.} Recently, signatures of self-organized criticality in ultracold Rydberg gas are demonstrated in Refs.~\cite{helmrich2020signatures, klocke2019controlling, PhysRevLett.126.123401, ding2020phase} (see also a brief discussion in~\cite{Rydberggassoc}). The dynamic processes can be described by a Langevin equation when the system exhibits an absorbing phase transition. A homogeneous ($D=0$, $\xi_t=0$) mean field to Langevin equations with active number density $n_a$ and total (active and ground/absorbing state, excluding auxiliary state) number density $n_{\rm tot}$ is given by~\cite{helmrich2020signatures}, 
\begin{equation}
\begin{split}
&\partial_t n_{a} =  (\kappa n_{a} - \tau) ( n_{\rm tot} - 2 n_{a}) - \gamma n_{a}, \\
&\partial_t n_{\rm tot} = - b \gamma n_{a},
\label{eq:langas}
\end{split}
\end{equation}
where $\kappa$, $\tau$ and $\gamma$ represent facilitation, spontaneous excitation and decay rates, respectively. The $b$ is a small parameter controlling the rate of decay to an auxiliary state. Throughout the process of SOC, final total number density always converge towards a fixed value $n_f$ over a wide range of initial inputs when exceed $n_f$. This indicates that $n_f$ is a critical point act as an attractor of the dynamics, showing that the final result is independent of initial number densities. Neglecting the small ingredient $\tau$ and taking a small $n_a$ instead of none at early time in the Eq.~\ref{eq:langas}, we derive a semi-analytic solution of $n_f$ in a stationary state,
\be
n_{f}=\frac{\gamma}{\kappa}\left(1-\frac{b}{2}\right), 
\label{eq:cnum}
\ee
which is well consistent with the solution to Eq.~\ref{eq:langas} at short time. On large time instead, it leads to a small deviation for $n_f$, as residual single atom excitation and subsequent loss occur with the rate $\tau$, contributing to an overall decay~\cite{helmrich2020signatures}.

Within the purview of the DM model, we employ the Eq.~\ref{eq:cnum} to ascertain the final relic abundance $Y_f$. The consideration of ingredient $\tau$ is rendered redundant owing to the stability of the DM. Note that a critical density, similar to the Eq.~\ref{eq:cnum} but without $b$, has been previously discussed in~\cite{klocke2019controlling, PhysRevLett.126.123401}. As we will show, if the parameter $b$ is comparable to one, it should not be neglected.


\,

{\it Self-organised criticality of dark matter.} In an analogy with the Rydberg gas, the DM $\chi$ and DP $\phi$ serve as ground and active states, respectively. Initially, both DM and DP possess nonzero abundances, which can be generated through the inflaton decay~\cite{Takahashi:2007tz}, gravitational production~\cite{Ren:2014mta, Garny:2015sjg, Mambrini:2021zpp, Kolb:2023ydq}, ultraviolet freeze-in~\cite{Moroi:1993mb, Bolz:2000fu} or decay of the false vacua~\cite{Asadi:2021pwo} in extremely early time of the Universe. It assumes that the initial abundance of DM is larger than the final relic abundance, while that of the DP is small. Subsequently, the DP grows {\it exponentially} in early time, similar to excitation avalanches in Rydberg gas. To achieve the irreversible decay of DP, we introduce an auxiliary field $\xi$ instead of the $\phi$ directly decaying to the SM particles, where the contribution of inverse decay can be negligible by assuming few initial abundance of the $\xi$. Suppose that DP has two different irreversible decay channels: one is the decay into an auxiliary field ($\xi$) with a rate $\Gamma_{\phi\to\xi}$, followed by the $\xi$ annihilates into the SM bath, and another one is the decay back into $\chi$ field at a rate $\Gamma_{\phi\to\chi}$, which is just to follow the Eq.~\ref{eq:langas} and can be neglect in subsequent analyses.

To quantitative analysis, the Boltzmann equations for the evolution of the DM and DP particles with 'time' $x=m_\chi/T$ and comoving number densites $Y=n/s$ are, 
\begin{eqnarray} \label{eq:boltz}
\frac{d Y_{\chi}}{d x} &=&-\frac{s(x)\langle \sigma_{\chi{\phi}\to\phi\phi} v \rangle}{x H(x)} \left( Y_\chi Y_{\phi} -  Y^2_{\phi} \frac{Y_\chi^{\rm eq}}{Y_\phi^{\rm eq}}\right) \nonumber\\
&&+\frac{\bar{\Gamma}_{\phi\to\chi} \,  Y_{\phi}}{x H(x)}, \,  \\
\frac{d Y_{\phi}}{d x} &=&\frac{s(x)\langle \sigma_{\chi{\phi}\to\phi\phi} v \rangle}{x H(x)}  \left( Y_\chi Y_{\phi}  -  Y^2_{\phi}\frac{Y_\chi^{\rm eq}}{Y_\phi^{\rm eq}}\right) \nonumber\\ 
                                 &&- \frac{(\bar{\Gamma}_{\phi \to \chi} + \bar{\Gamma}_{\phi \to \xi})  Y_{\phi}}{x H(x)},  \label{eq:boltz2}
\end{eqnarray}
where total number density is $Y_t=Y_\chi + Y_\phi$, corresponding to $d Y_t/d x= -\bar{\Gamma}_{\phi \to \xi} Y_\phi/(x H(x))$. $H(x)$ is the Hubble constant, $s(x)$ is the entropy density, and decay rate is $\bar{\Gamma}_{\phi \to \xi/\chi}=K_1(x) \Gamma_{\phi \to \xi/\chi}/K_2(x)$ with $K_n$ being the {\it n}th order modified Bessel function of the second kind. While the equilibrium abundance may in principle be different, throughout this analysis we assume $Y_\chi^{\rm eq}=Y_\phi^{\rm eq}$ at any time, since the mass difference between DM and DP is small. Suppose that $\langle \sigma_{\chi\phi\to\phi\phi} v \rangle = (\sigma v)_0 x^{k}$ with a constant $(\sigma v)_0$, where the exponent $k$ needs to be $k \geq2$ due to the exponential growth~\cite{Wintermantel2020EpidemicGA, Bringmann:2021tjr}. Note that other processes, such as $\phi \phi \leftrightarrow \chi \chi$ and $\phi \phi \to \xi\xi$, can be neglected.

\begin{figure}[t!]
\includegraphics[width=0.95\columnwidth]{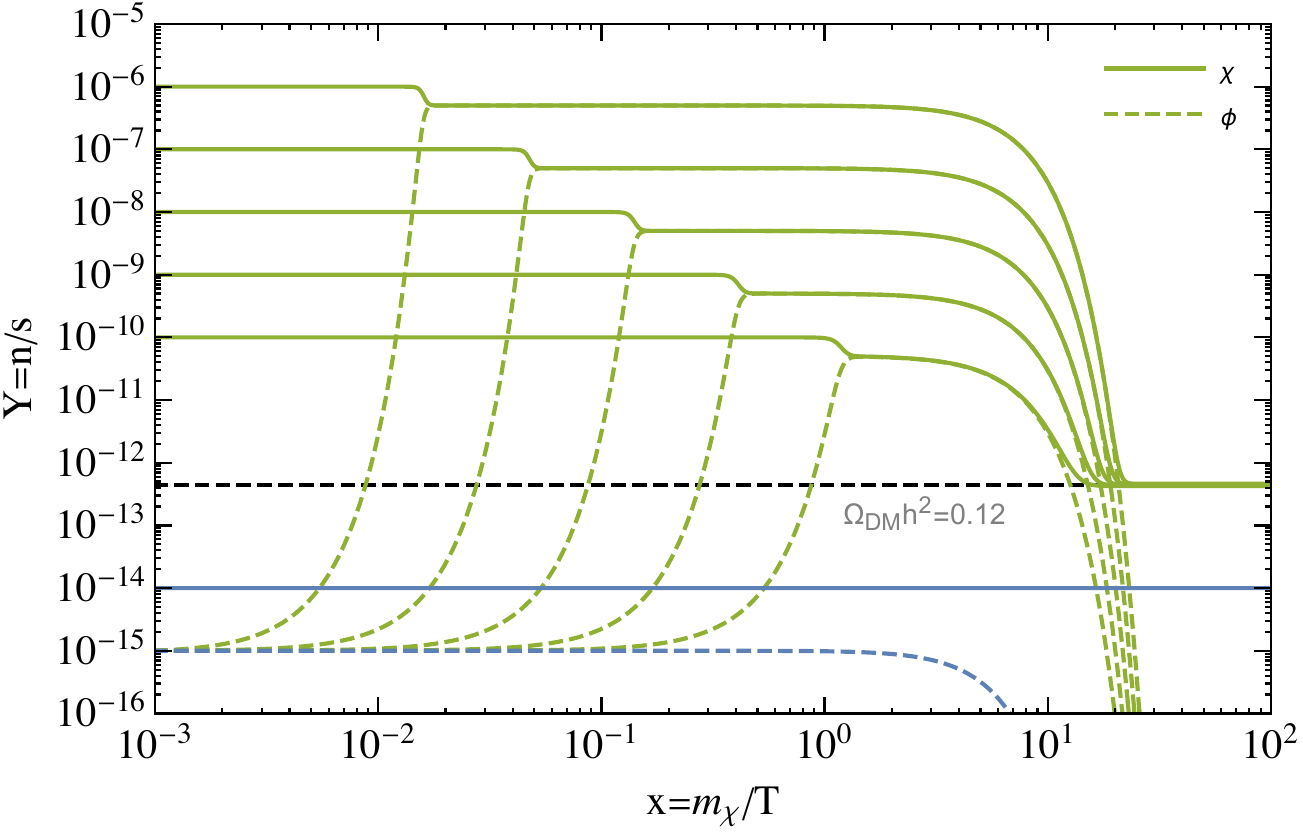}
\caption{The DM and DP abundances ($Y=n/s$) as a function of $x=m_\chi/T$, where $m_\chi=1$ TeV, $m_\phi=1.001 m_\chi$, $k=3$, and six distinct values ($10^{-14}, 10^{-10}, 10^{-9}, 10^{-8}, 10^{-7}, 10^{-6}$) of $Y_\chi^{\rm ini}$ and $Y_\phi^{\rm ini}=10^{-15}$. The color solid and dashed lines denote the DM and DP abundances, respectively. The black horizon dashed line describes the observed relic abundance with $\Omega h^2=0.12$.}
\label{fig:normalsoc}
\end{figure}
Obviously, the Eqs.~\ref{eq:boltz},~\ref{eq:boltz2} share similar forms with the Eq.~\ref{eq:langas} when drop the $\tau$ term. Thus the final abundance of DM can be written as 
\be
\label{eq:finrelic}
Y_f=\frac{\gamma_D}{\kappa_D}\left(1-\frac{\gamma_\xi}{2 \gamma_D} \right)
\ee 
where we define $\kappa_D\equiv s(x)$ $\langle \sigma_{\chi \phi\to\phi\phi} v \rangle/(x H(x))$, $\gamma_D\equiv \gamma_\xi + \gamma_\chi$ and $\gamma_{\xi/\chi}\equiv$ $\bar{\Gamma}_{\phi\to\xi/\chi}/(x H(x))$. The self-organizing dynamics are propelled by the interplay between parameters $\gamma_D$ and $\kappa_D Y_\phi$, in which both $\gamma_D$ and $\kappa_D$ are functions of $x$. It will nearly converge to a fixed point as long as the ratio $\gamma_D/\kappa_D$ remains independent of $x$. This condition is optimally satisfied when $k=3$ with large $m_\chi$, which is conducive to the establishment of a stable critical state. It is a well-established fact that the DM freeze-out occurs at temperature $x_f \sim \mathcal{O}(10)$, which provides a boundary condition that constrains the parameter $\gamma_D$ to be of order one at this time. As shown below, a number density equilibrium is established between the DM and DP. Therefore, we can roughly estimate the $\gamma_D$ by assuming that DM density begins to depart from equilibrium and freezes out instantly when $\kappa_D(x_f) Y_\phi(x_f)=1$, i.e. $n_\phi \langle \sigma_{\chi\phi\to\phi\phi} v \rangle= x_f\, H(x_f) $~\cite{Kramer:2020sbb, Frumkin:2022ror}. In the approximation where $Y_f \sim Y_f^{\rm min}= \gamma_D/(2\kappa_D)$, we have $Y_\phi(x_f)$ = $Y_\chi(x_f)$=$\gamma_D(x_f)/(2\kappa_D(x_f))$, thus obtain $\gamma_D(x_f)=2$. Suppose that the DM initial abundance $Y_\chi^{\rm ini} \gg Y_0$. The relation establishes a separated timescale: $\kappa_D(x_f) Y_\chi^{\rm ini} \gg \gamma_D(x_f)$, suggesting that the system exhibits rapid growth followed by slow decay, which is a key ingredient of the SOC system~\cite{Henkel2009NonEquilibriumPT,helmrich2020signatures,Grinstein1995,Bak1997HowNW}. Given that the parameter $\gamma_\chi(x_f)$ has no effect to a slow dissipation to boundary, it can be neglected. Consequently, Eq.~\ref{eq:finrelic} simplifies to $Y_f=\gamma_\xi/(2\kappa_D)$, implying $b=1$ in Eq.~\ref{eq:cnum}. The final abundance of DM can be simplified as, 
\be
Y_f=\frac{45\Gamma_{\phi \to \xi} K_1(x) x^{3-k}}{4\pi^2m_\chi^3 (\sigma v)_0 g_{s}(x)K_2(x)}
\label{eq:dmrelic}
\ee
where $g_s(x)$ is an effective number of entropy relativistic species of freedom. In Fig.~\ref{fig:normalsoc}, we show the evolution of DM and DP abundances by selecting several distinct initial values for the DM and a fixed value $Y_\phi^{\rm ini}=10^{-15}$ for the DP. In order for the $\gamma_\xi(x_f)$ to be $\mathcal{O}(1)$, we take $m_\chi=1\, {\rm TeV}$ and $\Gamma_{\phi \to \xi}=2\times10^{-13} \,{\rm GeV}^{-1}$. These choices correspond to $\gamma_\xi\approx 3.6$ at a reference value $x_f=25$, and $\kappa_D(x_f)$ is confined to $1.8 /Y_0$ at this juncture. This results in a weak interaction $\langle \sigma_{\chi \phi\to\phi\phi} v \rangle \approx 8 \times 10^{-8}\, {\rm GeV^{-2}}$. We demonstrate that initial inputs (green lines), significantly larger than a relic value $Y_0$, consistently approach the $Y_0$, whereas the input (blue line) below $Y_0$ never reach it.

\begin{figure}[t]
\includegraphics[width=0.95\columnwidth]{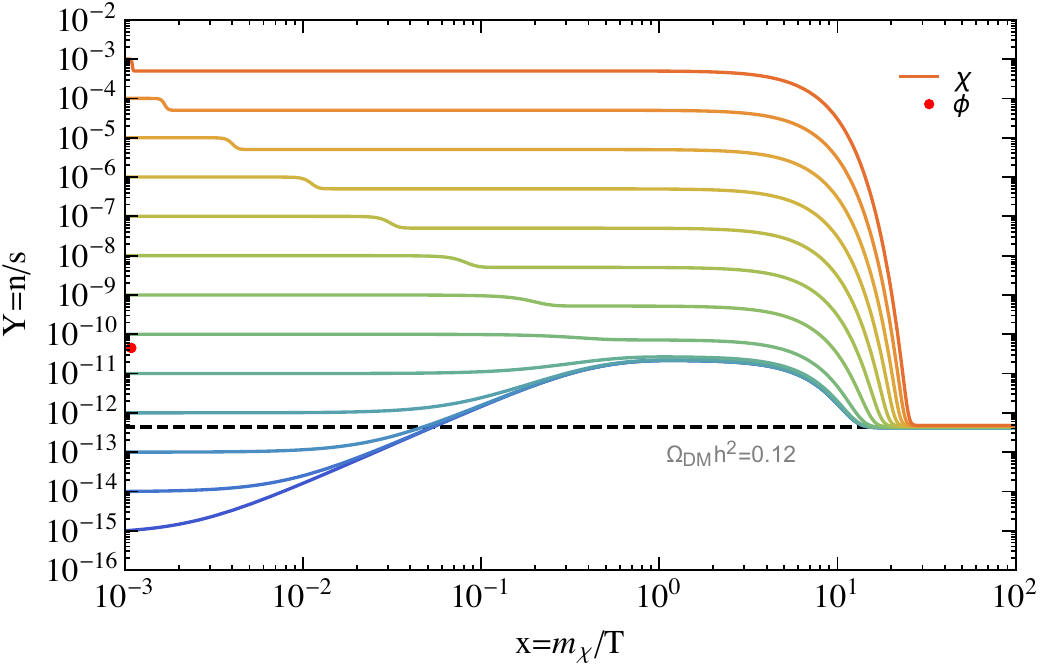}
\caption{Same parameters as in Fig.~\ref{fig:normalsoc} except initial abundances. For the sake of conciseness, only the evolution of the DM is displayed under an array of initial abundances within the range $[10^{-15}, 10^{-3}]$, while that of DP is maintained at a fixed value (red point) where $Y_\phi^{\rm ini}=10^2\,Y_0$.
}
\label{fig:randomsoc}
\end{figure}

Interestingly, we find that $\chi$ and $\phi$ rapidly achieve a number density equilibrium after exponential growth when one of them has an initial abundance much larger than $Y_0$. Therefore, a more general case emerges where the SOC process is triggered, provided that the initial total abundance satisfies $Y_t \gg Y_0$, even if the DM inputs is equal to or significantly less than the relic value. In Fig.~\ref{fig:randomsoc}, we illustrate a spectrum of initial abundances for $Y_\chi$, from tiny to thermal density, with a fixed value of the DP where $Y_\phi^{\rm ini}=10^2\, Y_0$ to ensure that the $Y_t$ is always much larger than $Y_0$. Furthermore, due to the presence of the equilibrium, the abundance of DM can be identified with that of the DP at the initial time. In other word, the SOC process can be emanate from a system that is in equilibrium. This suggests that the complex dynamics associated with SOC can originate from a thermal dark sector bath, and it highlights the system evolves towards criticality without the need for an external driving force that disrupts equilibrium.

It is worthwhile mentioning that the introduction of a decay channel for $\phi$ back into $\chi$ that would further increase the thermal cross section becomes a necessary element when the initial total abundance is comparable to its final one, while it needs the decay width $\Gamma_{\phi\to\chi} \gg \Gamma_{\phi\to\xi}$ at this stage. A beneficial aspect of this is that the SOC can be sustained under a less demanding condition, where it is sufficient for $Y_t>Y_0$, rather than a more stringent requirement $Y_t \gg Y_0$. However, to circumvent complexities and in accordance with the principle of Occam's razor, we have opted for a simplified model framework, which facilitates a clear elucidation of the underlying principles of the SOC system and aids in theoretical construction.

We find that it is possible for $m_\chi > m_\phi$ case, which is analogous to the zombie collision described in Ref.~\cite{ Berlin:2017ife}, however, it is essential to recognize that the interactions among $\chi$, $\phi$ and $\xi$ are explicitly distinct. In this case, the exponent $k$ is permitted to down to 2 and the mass difference between DM and DP may be less constrained. The suppression of $Y_\chi^{\rm eq}/Y_\phi^{\rm eq}$ at $x>1$ is counterbalanced by the residual $x$ present in the second term of the Boltzmann Equations, which renders the derivation of a general (semi-)analytical solution difficult. A relation $m_\phi=0.9 m_\chi$ is adaptable to the situation where $m_\chi=1$ TeV and $k=2$ and the relative mass for DP decreases with the reduction in the mass of DM.

In addition, in a multitude of models, the avalanches dynamics of SOC are found to adhere to scale invariance, which is exemplified by the power-law distributions of their magnitudes and durations. A $1/f$ noise is a well-known example of scale invariance in the SOC, which was initially proposed to explain spatial fractals and fractal time series in the BTW model~\cite{bak1987self}. The $1/f^\alpha$ form of power spectrum provides a more nuanced description of signal exhibiting scale invariance where the spectral exponent $\alpha$ is depend on numerical and/or experimental results, for example, the distribution of size of Rydberg excitations avalanche~\cite{helmrich2020signatures}, the power spectra of solar flare events and their interoccurance intervals~\cite{McAteer201525YO}, and displacement fluctuations of oscillators in the Ornstein-Uhlenbeck process~\cite{Kartvelishvili:2020thd}. It is imperative to recognize that $1/f$ noise is a few example of scale-invariant systems~\cite{Mandelbrot1982FractalGO}. In this works, our model is confined to a homogeneous fields in the absence of a slow regrowth or driving term, which is essential for the continuous occurrence of avalanches. This limitation impedes the examination of the $1/f$ noise relationship in the context of non-repetitive evolutionary process of DM.  A systematic study of the process of SOC is beyond the scope of this work, and we leave this to future works.
 
\,
{\it Dark matter phenomenology.}
We consider a simple phenomenological model, where $\chi,\,\phi$ and $\xi$ are real scalars, and $H$ is the SM Higgs with the interaction $\mathcal{L} \supset (\lambda_{\chi\phi}/3!) \chi \phi^3$+ $(\lambda_{\phi\xi}/3!)\phi \xi^3 + \lambda_{\xi h} \xi^2 |H|^2$. The decay constraint for DM are readily evaded through two intermediary fields. A straightforward scenario that accomplishes this is $m_\xi > m_\chi/5$ in the 5-body two-loop decay process. Given the instability of the $\phi$ and its production from non-thermal process, the DM mass can largely exceeds the unitary bound of $\mathcal{O}$(100) TeV~\cite{PhysRevLett.64.615}.

Besides, given that the parameter $\gamma_\xi$ is confined to $\mathcal{O}(1)$ and the freeze-out temperatures is fixed at $\mathcal{O}(10)$, it follows that the $\langle \sigma_{\chi\phi\to\phi\phi} v \rangle$ is restricted to a finite range ($10^{-9}\div 10^{ -6} \, {\rm GeV^{-2}}$) as well, especially the $\langle \sigma_{\chi\phi\to\phi\phi} v \rangle$ exhibits a negligible dependence on the large $m_\chi$ at fixed $\gamma_\xi$ and $x_f$ since $H(x) \propto m_\chi^2/x^2$ and $s(x) \propto m_\chi^3/x^3$. Numerically one obtains the $\langle \sigma_{\chi\phi\to\phi\phi} v \rangle \sim 6\times 10^{-7}\, {\rm GeV^{-2}}$ at $\gamma_\xi=10$, $x_f=40$ and $k=3$ when $m_\chi>1$ TeV. 

\,

{\it Conclusion.}
Within theoretical model of the DM, the evolution of DM is pivotal for understanding its intrinsic properties. In this scenario, we investigate the observed relic abundance of DM is independent of the initial inputs through semi-annihilation with heavier unstable DP, a process that is rooted in the dynamics of SOC. This exploration spans the parameter space between the freeze-out and freeze-in mechanisms, effectively bridging the gap between these two mechanisms. A significant outcome of this investigation is the realization that the dynamics of SOC can originate from a thermal dark bath, a novel perspective that has not been examined in previous researches. This approach could potentially offer deeper insights into the evolution of heavy DM in the early Universe.

\bigskip
\bigskip
{\bf Acknowledgements}  This work was supported in part by the National Science Foundation of China under the Grants No. 11925506, 12435004, and the Natural Science Foundation of Shandong province under the Grant No. ZR2022ZD26. 
\bigskip


\appendix

\bibliography{bibfile}

\begin{thebibliography}{98}
\bibitem [{Ryd()}]{Rydberggassoc}%
  \BibitemOpen
  \href@noop {} {\enquote {\bibinfo {title} {{In these experiments, the ground
  state of gases are constantly driven by a detuned laser field, resulting in
  rare and isolated Rydberg excitations at random positions. The excitations
  either spontaneously decay or trigger secondary excitations through
  facilitated excitation processes. Once the local density of ground states
  exceeds a threshold, which depends on facilitation and decay rate, the
  individual excitations will grow exponentially and trigger a spatially
  extended excitation avalanche characteristic of the SOC.}}}\ }\BibitemShut
  {NoStop}%
\end{thebibliography}

\end{document}